\documentclass{article}
\usepackage{spconf,amsmath,graphicx,hyperref}
\usepackage{amssymb}
\usepackage{xcolor}
\usepackage{booktabs}
\usepackage{threeparttable}

\title{Lightweight Implicit Neural Network for Binaural Audio Synthesis}
\name{
    Xikun Lu$^{1}$ \qquad
    Fang Liu$^{2}$ \qquad
    Weizhi Shi$^{3}$ \qquad
    Jinqiu Sang$^{2}$\textsuperscript{*}
        \thanks{* Corresponding author: Jinqiu Sang (jqsang@mail.ecnu.edu.cn)}
}
\address{%
  $^{1}$ Shanghai Institute of Artificial Intelligence for Education, East China Normal University, China\\
  $^{2}$ School of Computer Science and Technology, East China Normal University, China\\
  $^{3}$ Department of Big Data and Information Engineering, Guizhou Industry Polytechnic College, China
}

\begin{document}
\ninept
\maketitle
\begin{abstract}
High-fidelity binaural audio synthesis is crucial for immersive listening, but existing methods require extensive computational resources, limiting their edge-device application. To address this, we propose the Lightweight Implicit Neural Network (Lite-INN), a novel two-stage framework. Lite-INN first generates initial estimates using a time-domain warping, which is then refined by an Implicit Binaural Corrector (IBC) module. IBC is an implicit neural network that predicts amplitude and phase corrections directly, resulting in a highly compact model architecture. Experimental results show that Lite-INN achieves statistically comparable perceptual quality to the best-performing baseline model while significantly improving computational efficiency. Compared to the previous state-of-the-art method (NFS), Lite-INN achieves a 72.7\% reduction in parameters and requires significantly fewer compute operations (MACs). This demonstrates that our approach effectively addresses the trade-off between synthesis quality and computational efficiency, providing a new solution for high-fidelity edge-device spatial audio applications.
\end{abstract}
\begin{keywords}
Binaural audio synthesis, implicit neural network, lightweight model, edge-device application.
\end{keywords}
\section{Introduction}
\label{sec:intro}

Spatial audio is fundamental to building immersive multimedia experiences and crucial for creating a sense of realism and presence \cite{broderick2018importance}. Among numerous rendering methods, binaural audio synthesis stands out as a notable approach. By simulating the acoustic path that sound takes to reach the human ear, it can reproduce a three-dimensional sound field using headphones. Consequently, it has been widely used in virtual reality (VR), augmented reality (AR), and interactive entertainment \cite{serafin2018sonic,madmoni2018direction}.

Traditional binaural audio synthesis relies primarily on head-related transfer functions (HRTFs), which are based on physical measurements. HRTFs accurately describe the linear filtering effects of sound waves as they propagate, influenced by the listener's physiological characteristics \cite{xie2013head}. While effective, this approach typically relies on a convolutional static HRTF database, which poses challenges when rendering continuous dynamic scenes \cite{li2020measurement}. Furthermore, its inherent linear time-invariant assumption limits its ability to model complex nonlinear acoustic phenomena in the real world \cite{brinkmann2017authenticity}.

With the increasing maturity of deep learning technologies \cite{li2019neural, michelsanti2021overview}, researchers have developed a variety of end-to-end binaural audio synthesis methods \cite{gebru2021implicit,richard2021neural,huang2022end,NEURIPS2022_95f03faf,lee2023neural,he2025dual,zhang2025two}. These methods aim to directly learn mappings from source audio and spatial metadata to the binaural output. Some advanced approaches even integrate visual, textual, and other information to create multimodal systems \cite{lu2025deep, NEURIPS2018_01161aaa,li2024tas,heydari2025immersediffusion}. Although these multimodal methods enhance spatial perception, they generally increase computational complexity and rely on the availability of multimodal data. Powerful and efficient pure audio synthesis models remain a fundamental prerequisite for many applications. However, whether directly generating waveforms in the time domain \cite{richard2021neural,NEURIPS2022_95f03faf} or predicting spectra in the frequency domain \cite{lee2023neural,he2025dual,zhang2025two}, both aim to capture the complex details of acoustic propagation accurately. However, to achieve high fidelity, they generally rely on large-scale neural network architectures. This reliance on model capacity results in significant memory and computational requirements, severely hindering their deployment on resource-constrained edge-devices. Therefore, developing efficient and lightweight alternatives is crucial for promoting the practical application of binaural audio technology.

The emergence of Implicit Neural Representations (INRs) \cite{xie2022neural,NEURIPS2020_53c04118} provides new inspiration for efficient signal modeling. Instead of storing discrete data in large networks, INRs leverage compact networks to learn a continuous function that maps input coordinates to signals. This approach has achieved remarkable success in a wide range of applications, from neural radiance fields in vision \cite{NEURIPS2020_53c04118} to their acoustic counterpart, neural sound fields~\cite{NEURIPS2022_151f4dfc,NEURIPS2023_760dff0f}. In the audio field, INRs have been applied to tasks such as sound field reconstruction \cite{NEURIPS2022_35d5ad98}, room impulse response modeling \cite{liang2023neural,brunetto2025neraf}, and even direct HRTF synthesis \cite{zhang2023hrtf,lu2025bicg}. These models can encode complex acoustic details in low-parameter networks, providing strong motivation for their application to the efficiency challenge of binaural audio synthesis.
Inspired by this, we propose the \textbf{L}ightweight \textbf{I}mplicit \textbf{N}eural \textbf{N}etwork (Lite-INN). Its core innovation lies in an Implicit Binaural Corrector (IBC) that takes spatial pose, ear index, frequency, and time encoding as continuous inputs. It predicts logarithmic amplitude and phase corrections via a small multilayer perceptron (MLP). This network is capable of efficiently modeling complex acoustic transfer functions at a low computational cost. Our contributions are summarized as follows:

1. We propose a new paradigm for modeling spectral correction as a continuous implicit function, implemented via a novel IBC.

2. Based on this paradigm, we construct a lightweight end-to-end network (Lite-INN) with significantly fewer parameters, making it suitable for edge-device deployment.

3. Experiments show that Lite-INN achieves higher computational efficiency while maintaining comparable synthesis quality to previous work. Our source code and audio samples are available online\footnote{\url{https://github.com/Luxikun669/Lite-INN}}.

\begin{figure*}
    \centering
    \includegraphics[width=0.9\linewidth]{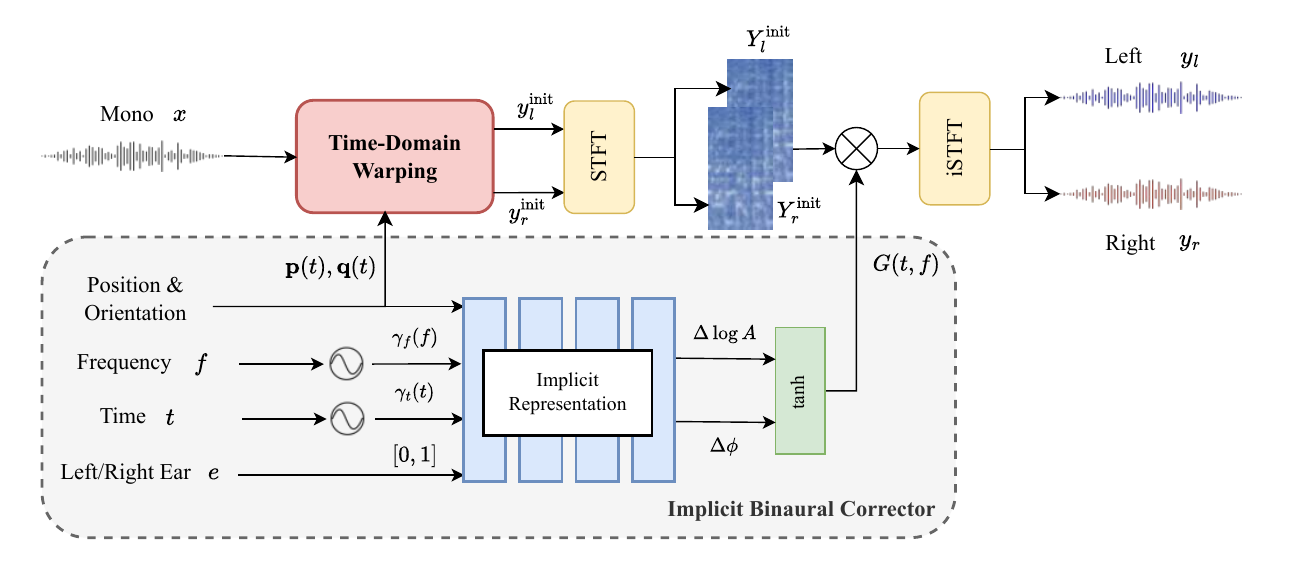}
    \caption{The proposed Lite-INN architecture, a two-stage process combining a Time-Domain Warping (TDW) network for initial synthesis with an Implicit Binaural Corrector (IBC) for spectral refinement.}
    \label{fig1}
\end{figure*}

\section{Proposed Method}
\label{sec:format}

\subsection{Architecture Overview}

The goal of this work is to synthesize a binaural audio signal $y \in \mathbb{R}^{B \times 2 \times L}$ from a monaural input $x \in \mathbb{R}^{B \times 1 \times L}$. This synthesis is conditioned on a time-varying pose $P(t) \in \mathbb{R}^{B \times 7 \times T } $, which specifies the position $\mathbf{p}(t)\in \mathbb{R}^{3}$ and orientation $\mathbf{q}(t)\in \mathbb{R}^{4}$ of the sound source relative to the listener. 

As shown in Fig.~\ref{fig1}, the Lite-INN architecture consists of two-stage. The first stage uses a Time-Domain Warping (TDW) module~\cite{richard2021neural} for initial synthesis. This module resamples the monaural input based on the geometric information specified by the position $\mathbf{p}(t)$ and orientation $\mathbf{q}(t)$ to approximate the Interaural Time Difference (ITD), producing the intermediate signals $(y_{l}^{\text{init}},y_{r}^{\text{init}})$. In the second stage, the intermediate signals are converted to their complex spectrogram representation via a Short-Time Fourier Transform (STFT). The IBC then operates in this time-frequency domain, predicting a complex-valued gain mask $G(t,f)$ and applying fine-grained spectral corrections. The final binaural audio is obtained by performing element-wise complex multiplication of this gain, followed by an inverse STFT (iSTFT). This entire forward process is formulated as follows:
\begin{equation}
\label{eq:main_formula}
y = \text{iSTFT}(\text{STFT}(\text{TDW}(x | \mathbf{p}(t),\mathbf{q}(t)) \odot G(t,f))
\end{equation}

\begin{table*}[t]
\centering
\begin{threeparttable}
\caption{Quantitative results comparison with state-of-the-art models on the Binaural Speech dataset.}
\label{tab:main_results}
\begin{tabular*}{\textwidth}{@{\extracolsep{\fill}} l c c c c c c c @{}}
\toprule
\textbf{Model} & \textbf{Year} & \textbf{\# Param. (M)} & \textbf{MACs (G) $\downarrow$} & \textbf{Wave-$\ell_2$ $\downarrow$} & \textbf{Amplitude-$\ell_2$ $\downarrow$} & \textbf{Phase-$\ell_2$ $\downarrow$} & \textbf{IPD-$\ell_2$ $\downarrow$} \\
\midrule
WaveNet~\cite{van2016wavenet}        & arXiv'16 & 4.65 & 22.34 & 0.179 & 0.037 & 0.968 & 1.114 \\
WarpNet~\cite{richard2021neural}        & ICLR'21 & 8.59 & 19.15 & 0.167 & 0.048 & 0.807 & 1.166 \\
WarpNet$^\ast$~\cite{NEURIPS2022_95f03faf} & NeurIPS'22 & -- & -- & 0.157 & 0.038 & 0.838 & -- \\
BinauralGrad~\cite{NEURIPS2022_95f03faf}   & NeurIPS'22 & 13.8 & 229.4 & \textbf{0.128} & \textbf{0.030} & 0.837 & 1.099 \\
NFS~\cite{lee2023neural}            & ICASSP'23 & 0.55 & 3.400 & 0.172 & 0.035 & 0.999 & 1.250 \\
DPATFNet~\cite{he2025dual}      & ICASSP'25 & 2.42 & 15.64 & 0.148 & 0.037 & \textbf{0.717} & \textbf{1.020} \\
\textbf{Lite-INN (Ours)} & --  & \textbf{0.15} & \textbf{2.670} & 0.167 & 0.040 & 0.857 & 1.233 \\
\bottomrule
\end{tabular*}


\end{threeparttable}
\end{table*}

\subsection{Implicit Binaural Corrector}
\label{sec:ibc}

The core of our method is the IBC module, which is an implicit representation that models a continuous function mapping spatio-temporal-frequency coordinates to corresponding spectral corrections. 
For each time-frequency bin $(t,f)$, a high-dimensional coordinate vector $\textbf{c}$ is constructed as input to the network. This vector consists of the concatenation of the source pose $(\mathbf{p},\mathbf{q})$ and the one-hot vector of the ear index $e$. To explicitly embed the time-frequency coordinates, the discrete indices $f$ and $t$ are converted to a continuous vector representation using a sinusoidal positional encoding. The Frequency Positional Encoding (FreqPE, $\gamma_f(f)$) is defined as:
\begin{equation}
\begin{split}
    \gamma_f(f) =
    \big(\sin(2^0\!\cdot\!2\pi f),\, \cos(2^0\!\cdot\!2\pi f),\, \ldots,\, \\
      \sin(2^{N_f-1}\!\cdot\!2\pi f),\, \cos(2^{N_f-1}\!\cdot\!2\pi f)\big)
\end{split}
\end{equation}
where $N_f$ is the number of frequency encoding bands. An analogous Time Positional Encoding (TimePE, $\gamma_t(t)$) is computed for the time index $t$ using $N_t$ bands. The complete coordinate vector $\textbf{c}$, which concatenates the pose, one-hot ear index, FreqPE, and TimePE, is then fed into the core of the IBC.

The core is an MLP consisting of $L$ hidden layers, each containing $H$ nodes and a SiLU activation function. The final linear layer of the MLP outputs an unconstrained two-dimensional correction vector $(\Delta\log A, \Delta\phi)$. To ensure training stability and limit the range of the correction, these raw outputs are scaled using the hyperbolic tangent (tanh) function, resulting in the final adjustment values:
\begin{equation}
\delta_A(t,f) = \alpha\,\tanh\!\big(\Delta\log A(t,f)\big)
\end{equation}
\begin{equation}
\delta_\phi(t,f)   = \pi\,\tanh\!\big(\Delta\phi(t,f)\big)
\end{equation}
with a fixed scale $\alpha=0.8$. Finally, these scaled adjustment values are used to construct the complex gain $G(t,f)$ for a specific time-frequency bin, which precisely modulates the amplitude and phase:
\begin{equation}
G(t,f) = \exp(\delta_A(t,f)) \cdot \exp(j \cdot \delta_\phi(t,f))
\end{equation}

By querying each coordinate in the target spectrogram, we can efficiently generate the complete correction mask required for high-fidelity binaural audio synthesis.

\subsection{Loss Function}
\label{sec:loss}

To train the network, we use a loss function that considers both waveform fidelity and phase accuracy. The total loss $\mathcal{L}$ is a weighted sum of the time-domain $\ell_2$ loss and the frequency-domain phase loss, adapted from~\cite{he2025dual}:
\begin{equation}
\mathcal{L}(y, y^*)
= \lambda_{\text{1}}\,
\underbrace{\big\|\,y - y^*\,\big\|_{2}}_{\ell_{2}}
\;+\;
\lambda_{\text{2}}\,
\underbrace{\big\|\,\angle Y-\angle Y^{*}\,\big\|_{1}}_{\mathcal{L}_{\text{phase}}}
\end{equation}
where $y$ is the predicted waveform (from Eq.~(\ref{eq:main_formula})),
$y^*$ is the ground-truth waveform,
$Y$ and $Y^*$ are the corresponding complex spectrograms,
$\angle(\cdot)$ extracts the element-wise phase in $(-\pi,\pi]$,
and $\lambda_{\text{1}}$ and $\lambda_{\text{2}}$ are loss weights.
The phase term penalizes the absolute phase difference, which is crucial for accurate spatial localization~\cite{richard2021neural}.

\section{Experiments}
\label{sec:pagestyle}

\subsection{Datasets}

We use the Binaural Speech dataset \cite{richard2021neural} for all training and testing procedures. The dataset contains approximately two hours of 48 kHz audio recordings from eight speakers. For each monaural audio input, the dataset provides the corresponding ground-truth binaural signal and the time-aligned 6-DoF pose of the sound source, recorded at 120 Hz. We use the official data split provided by the dataset to spilt the training, validation, and test sets.

\subsection{Baselines}

The performance of Lite-INN is benchmarked against several existing models. We include WaveNet~\cite{van2016wavenet} as a foundational autoregressive baseline. Our comparison also includes WarpNet~\cite{richard2021neural}, which serves as the foundation for our model's first stage, and its reimplementation WarpNet*~\cite{NEURIPS2022_95f03faf}. To benchmark against high-fidelity generative approaches, we include BinauralGrad~\cite{NEURIPS2022_95f03faf}, a diffusion-based model representing the state-of-the-art in audio quality. The comparison extends to frequency-domain methods, including NFS~\cite{lee2023neural}, which predicts spectral shifts, and DPATFNet~\cite{he2025dual}, a model that employs attention mechanisms on time-frequency representations.

\subsection{Implementations Details}

For signal processing, we compute the STFT using a 512-sample Hamming window with a hop length of 256 samples. The TDW module is adapted from~\cite{richard2021neural}, but with one fewer convolutional layer to reduce complexity. The IBC module is an MLP with 3 hidden layers and 256 units per layer. The coordinate input to the IBC is constructed from the source pose (position and orientation), ear index, FreqPE, and TimePE. The FreqPE uses 8 bands, and the TimePE uses 12 bands. The model is trained for 100 epochs with a batch size of 32. We use the AdamW optimizer and a Cosine Annealing strategy, where the learning rate decays from $1 \times 10^{-3}$ to $1 \times 10^{-6}$. The loss weights $\lambda_{\text{1}} $ and $\lambda_{\text{2}}$ are set to 1.0 and 0.01, respectively.

\subsection{Evaluation Metrics}

We assess model performance through a combination of quantitative and perceptual evaluations. For quantitative analysis, we compute the Mean Squared Error (MSE) between the synthesized and ground-truth audio. This includes Wave-$\ell_2$ for fidelity on the time-domain waveform, along with Amplitude-$\ell_2$ and Phase-$\ell_2$, which measure error on the magnitude and phase of the STFT spectrograms. To specifically evaluate the rendering of spatial cues, we also report IPD-$\ell_2$, the MSE on the Interaural Phase Difference \cite{li2024diffbas}. For perceptual analysis, we conduct Mean Opinion Score (MOS) listening tests. Listeners rated the audio based on three aspects: the overall naturalness and quality (MOS-Q), the perceived accuracy of the sound source's spatialization (MOS-S), and the fidelity to the ground-truth reference (MOS-Sim).

\begin{table}[t]
\centering
\caption{Ablation experiment results of Lite-INN.}
\label{tab:abl}
\setlength{\tabcolsep}{3.5pt}
\renewcommand{\arraystretch}{1.05}
\begin{tabular*}{\columnwidth}{@{\extracolsep{\fill}} l c c c c}
\toprule
\textbf{Model} & \textbf{Wave-$\ell_2$ $\downarrow$} & \textbf{Amp-$\ell_2$ $\downarrow$} & \textbf{Phase-$\ell_2$ $\downarrow$} & \textbf{IPD-$\ell_2$ $\downarrow$} \\
\midrule
\textbf{Lite-INN (Ours)} & \textbf{0.167} & \textbf{0.040} & \textbf{0.857} & \textbf{1.233} \\
w/o TDW              & 0.329 & 0.051 & 1.345 & 1.666 \\
w/o IBC              & 0.377 & 0.058 & 1.038 & 1.461 \\
w/o FreqPE           & 0.228 & 0.044 & 0.975 & 1.417 \\
w/o TimePE           & 0.168 & \textbf{0.040} & 0.864 & 1.325 \\
\bottomrule
\end{tabular*}
\end{table}

\begin{figure*}
    \centering
    \includegraphics[width=1.0\linewidth]{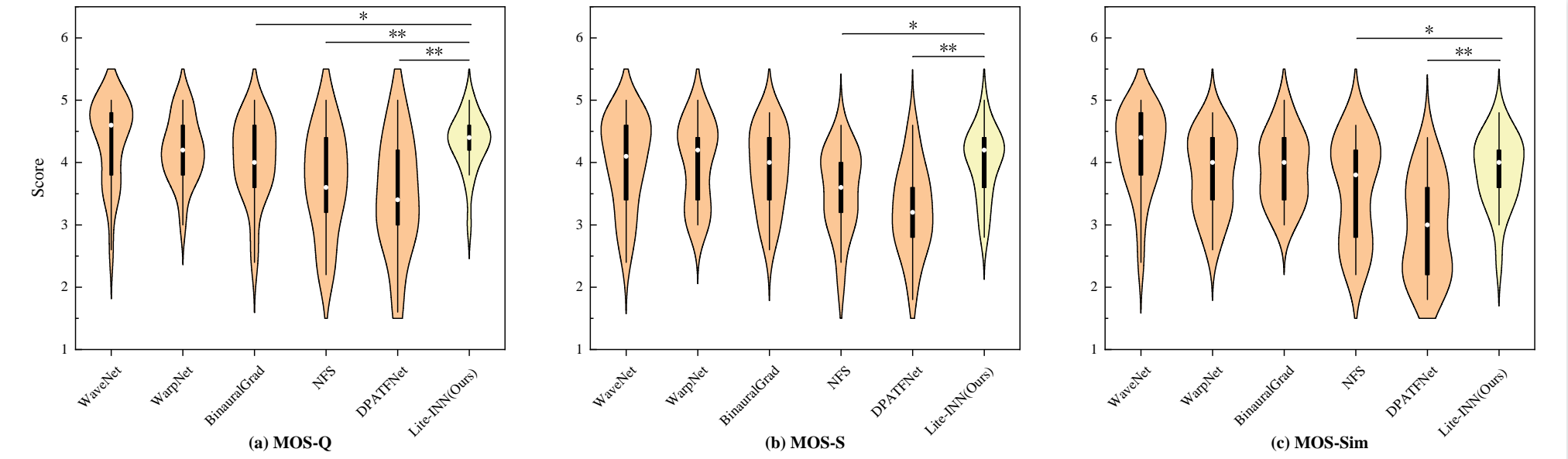}
    \caption{Violin plots of the MOS listening tests for (a) MOS-Q, (b) MOS-S, and (c) MOS-Sim. Statistical significance was determined by pairwise Wilcoxon Signed-Rank Tests comparing our model (Lite-INN) to each baseline (* indicates $p < 0.05$, ** indicates $p < 0.001$).}
    \label{fig:placeholder}
\end{figure*}

\section{Results and Discussion}
\label{sec:typestyle}

\subsection{Quantitative and Perceptual Evaluation}

The quantitative and perceptual results are presented in Table~\ref{tab:main_results} and Fig.~\ref{fig:placeholder}, respectively. First, we analyze the quantitative results from Table~\ref{tab:main_results}. The data shows that while Lite-INN does not secure the top performance in any individual objective metric, it maintains a competitive performance profile against baseline models. Notably, its objective scores are achieved with unparalleled efficiency. With only 0.15 M parameters and 2.67 GMACs, Lite-INN outperforms the most lightweight model. Compared to NFS, the most compact baseline, Lite-INN achieves superior results on three of the four metrics with 72.7\% fewer parameters. This result supports the primary goal of our work: achieving a balance between objective performance and computational feasibility, making it highly suitable for edge-device applications. The model achieves a superior real-time factor (RTF)\footnote{The RTF is measured with Intel(R) Xeon(R) Gold 6146 CPU.} of 0.121, which directly demonstrates its real-time processing capability and fully demonstrates its applicability.

Next, we turn to the perceptual evaluation, visualized in Fig.~\ref{fig:placeholder}. For these tests, 21 participants rated the audio on a 1-to-5 scale for MOS-Q, MOS-S, and MOS-Sim. The results indicate that WaveNet, a classic autoregressive model, achieved the highest mean scores, with our Lite-INN model ranking second across all three categories. In addition, the critical finding comes from the statistical analysis. A Wilcoxon Signed-Rank Test reveals no statistically significant difference ($p > 0.05$) between the perceptual scores of Lite-INN and WaveNet. In contrast, Lite-INN shows a statistically significant improvement over other contemporary baselines such as DPATFNet and NFS ($p < 0.05$).

This work's central conclusion is that Lite-INN successfully resolves the trade-off between quality and efficiency. It delivers perceptual quality statistically on par with WaveNet, the top-performing model, while reducing parameters by 96.8\% and MACs by 88.0\%. The discrepancy between Lite-INN's objective scores and its top-tier subjective performance suggests our method excels at minimizing perceptually relevant artifacts not fully captured by traditional $\ell_2$-based metrics.

\subsection{Interpretability Analysis}



\begin{figure}
    \centering
    \includegraphics[width=1\linewidth]{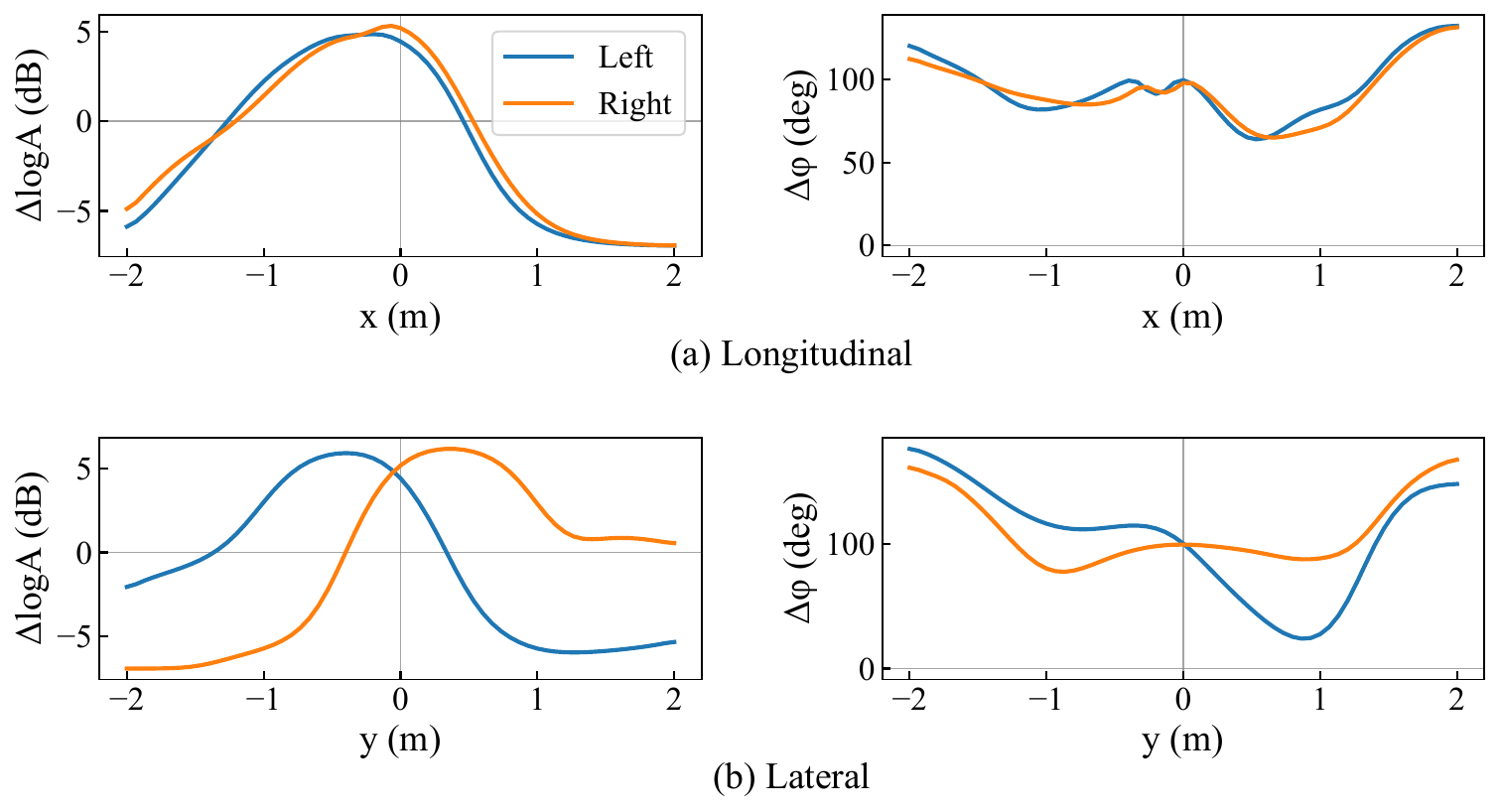}
    \caption{Plot of $\Delta\log A$ and  $\Delta\phi$ averaged over the covering frequencies for the channel with the most dominant intensity.}
    \label{fig3}
\end{figure}

To explain the behavior learned by the IBC, we visualized its predicted magnitude ($\Delta\log A$) and phase ($\Delta\phi$) corrections against the sound source's position, as shown in Fig.~\ref{fig3}. For lateral motion (left-right), the model predicts opposite amplitude corrections for both ears, with the gain of the ipsilateral ear increasing while the gain of the contralateral ear decreases. This behavior accurately simulates the interaural level difference (ILD), a primary localization cue derived from the head shadow effect. Meanwhile, the phase correction exhibits an antisymmetric relationship, diverging as the source moves away from the center, corresponding to the interaural time difference (ITD).

In contrast, for longitudinal motion (front-back) along the median plane, the corrections for both ears remain nearly symmetric. This observation is consistent with the physical principle that sources located in this plane produce negligible ILD and ITD. Overall, these visualizations demonstrate that the IBC has learned a functionally meaningful and generalizable representation of the underlying acoustic transfer function.

\subsection{Ablation Study}

The ablation study in Table~\ref{tab:abl} validates our architectural design. The necessity of our two-stage approach is confirmed by the sharp performance drop either without the TDW stage (w/o TDW) or without the IBC (w/o IBC). These components are symbiotic: the TDW provides a robust initial synthesis, after which the IBC precisely refines in the spectral domain. The role of positional encodings is also clarified. Disabling frequency encoding (w/o FreqPE) significantly degrades performance, confirming that providing explicit frequency coordinates is critical for the implicit network to model frequency-dependent spatial cues. In contrast, the minimal impact of removing time encoding (w/o TimePE) suggests the model's temporal awareness is sufficiently handled by the TDW stage, allowing the IBC to focus on the spectral task.

\section{Conclusion}
\label{sec:illust}

In this paper, we propose a lightweight binaural audio synthesis framework, named Lite-INN. Our method utilizes a two-stage process where an initial warped signal is refined by an Implicit Binaural Corrector (IBC) module, which models spectral corrections as a continuous function of spatio-temporal-frequency coordinates. Our experimental results demonstrate that Lite-INN achieves statistically comparable subjective listening quality to baseline models while maintaining superior objective metrics.  
Notably, Lite-INN operates with superior computational efficiency, reducing parameters by 72.7\% and MACs by 21.5\%  compared to the most compact baseline. This validates the potential of implicit neural representations for enabling real-time, high-fidelity spatial audio rendering on resource-constrained devices.

\bibliographystyle{IEEEbib}
\bibliography{strings,refs}

\end{document}